\documentclass[aps,10pt,pre,floatfix,twocolumn,superscriptaddress]{revtex4-1}
\usepackage{color}
\usepackage{graphicx}
\usepackage{subfigure,amsmath, amsthm, amssymb}

\newcommand{\be}{\begin{equation}}
\newcommand{\ee}{\end{equation}}
\newcommand{\bea}{\begin{eqnarray}}
\newcommand{\eea}{\end{eqnarray}}

\begin{document}
\title{Phase transitions in a system of hard  $Y$-shaped particles on the triangular lattice}
\author{Dipanjan Mandal}
\email{mdipanjan@imsc.res.in}
\affiliation{The Institute of Mathematical Sciences, C.I.T. Campus, Taramani, Chennai 600113, India}
\affiliation{Homi Bhabha National Institute, Training School Complex, Anushakti Nagar, Mumbai 400094, India}
\author{Trisha Nath}
\email{trisha.nath@theorie.physik.uni-goettingen.de}
\affiliation{Institut f\"{u}r Theoretische Physik, Georg-August-Universit\"{a}t G\"{o}ttingen, 37077 G\"{o}ttingen, Germany}
\author{R. Rajesh} 
\email{rrajesh@imsc.res.in}
\affiliation{The Institute of Mathematical Sciences, C.I.T. Campus, Taramani, Chennai 600113, India}
\affiliation{Homi Bhabha National Institute, Training School Complex, Anushakti Nagar, Mumbai 400094, India}

\date{\today}
\begin{abstract}
We study the different phases and the phase transitions  in a system of $Y$-shaped particles, examples 
of  which include Immunoglobulin-G and trinaphthylene molecules,
on a triangular lattice interacting exclusively  through excluded volume interactions. Each particle consists
of a central site and three of its six nearest neighbours chosen alternately, such that there are two types
of particles which are mirror images of each other. We study the equilibrium properties of the system using
grand canonical Monte Carlo simulations that implements an algorithm with cluster moves that is able to
equilibrate the system at densities close to full packing. We show that, with increasing density, the system undergoes
two entropy-driven phase transitions with two broken-symmetry phases. At low densities, the system is in a disordered phase. 
As intermediate phases, there is a solid-like sublattice phase in which one type of particle is preferred over the other and the particles 
preferentially occupy one of four sublattices, thus breaking both particle-symmetry as well as translational invariance. At even
higher densities, the phase is a columnar phase, where the particle-symmetry is restored, and the particles preferentially
occupy even or odd rows along one of the three directions. This phase has translational order in only one direction,
and breaks rotational invariance. From finite size scaling, we demonstrate that both the transitions are first order in nature. We also
show that the simpler system with only one type of particles undergoes a single discontinuous phase transition from a disordered
phase to a solid-like sublattice phase with increasing  density of particles.
\end{abstract}
\maketitle

\section{Introduction}
The study of the phases and critical behavior of lattice systems of hard particles having different geometrical shapes
has been of continued interest in  classical statistical mechanics, not only from the point of view
of how complex phases arise from simple interactions, but also for understanding 
how different universality classes of continuous phase transitions depend on the shape of the particles. 
Such hard core lattice gas (HCLG) models have also been of interest in the context of the  freezing 
transition~\cite{1962-aw-pr-phase,1957-aw-jcp-phase}, directed and undirected lattice animals
~\cite{1982-d-prl-equivalence,1983-d-prl-exact,2003-bi-jsp-dimensional},
the Yang-Lee edge singularity~\cite{1981-ps-prl-critical}, and in absorption of molecules onto substrates
~\cite{1985-twpbe-prb-two,2000-psb-ssr-phase,2001-mbr-ss-static,1998-k-jec-lattice,1985-bkuoabe-prl-phase}. 
Since only excluded volume interactions are present, temperature plays no role, and phase transitions, if any,
are entropy driven. Many different shapes have been studied in the literature. Examples include
triangles~\cite{1999-vn-prl-triangular}, 
squares~\cite{1967-bn-jcp-phase,1966-bn-prl-phase,1966-rc-jcp-phase,2012-rd-pre-high,2016-ndr-epl-stability,2017-mnr-jsm-estimating},
dimers~\cite{1961-k-physica-statistics,1961-tf-pm-dimer,2003-hkms-prl-coulomb,2017-naq-arxiv-polyomino},
mixture of squares and dimers~\cite{2015-rdd-prl-columnar,2017-mr-pre-columnar},
Y-shaped particles~\cite{C3RA45342A,2015-rthrg-tsf-impact},  
tetrominoes~\cite{2002-mhs-jcp-simple,2009-bsg-langmuir-structure},
rods~\cite{2007-gd-epl-on,2013-krds-pre-nematic,2017-gkao-pre-isotropic,2017-vdr-arxiv-different}, 
rectangles~\cite{2014-kr-pre-phase,2015-kr-pre-asymptotic,2015-nkr-jsp-high,2017-gvgmv-jcp-ordering},
discs~\cite{2007-fal-jcp-monte,2014-nr-pre-multiple}, and 
hexagons~\cite{1980-b-jpa-exact}. The hard hexagon model on the triangular lattice is the only solvable
model.

In this paper, we focus on hard $Y$-shaped particles on the triangular lattice. Particles with this shape arise in 
different contexts. A well known example is Immunoglobulin-G (IgG), an antibody present in human blood,
consisting of four peptide chains, two identical heavy chains and two identical light chains~\cite{Immunobiology}.
IgG has many therapeutic usages and study of different phases of $Y$-shaped 
particles~\cite{C3RA45342A,2008-lrj-jcp-polymer,2015-rthrg-tsf-impact} is important to understand the effect of density on the viscosity of the liquid.
Another example of a $Y$-shaped particle that is relevant for applications
is trinaphthylene. It has been useful 
to create a NOR logic gate on Au(111) surface~\cite{soe,godlewski}, in which napthylene branches of the molecule comes in 
contact of Au atom and act as input of logic gate.

Motivated by these applications, there have been a few numerical studies~\cite{2015-rthrg-tsf-impact,C3RA45342A} of systems of  
$Y$-shaped particles on a triangular lattice.
Each particle constitutes of a central site and three of its nearest neighbours chosen alternately. There are two
types of particles possible depending on which of the neighbours are chosen. In Refs.~\cite{2015-rthrg-tsf-impact,C3RA45342A}, in
addition to the hard core constraint, there are additional attractive interactions between 
the arms of neighbouring particles.
At low temperatures, a single first order phase transition from a disordered phase to a high-density ordered phase was observed.
The high density phase consists of mostly only one of the two types of $Y$-shaped particles, and has a solid-like sublattice order. 
For temperatures above a critical temperature, there are no density-driven phase transitions~\cite{2015-rthrg-tsf-impact,C3RA45342A}. 
At the critical temperature, the transition has been argued to belong to the
Ising universality class~\cite{2015-rthrg-tsf-impact}. For $Y$-shaped particles with larger arm lengths, other phases are
also seen~\cite{C3RA45342A}.

In this paper, we determine the different phases and nature of the phase transitions when only
excluded volume interactions are present, corresponding to the infinite temperature limit of the model studied in 
Refs.~\cite{2015-rthrg-tsf-impact,C3RA45342A}. 
We show that the sublattice phase  at high densities which breaks particle symmetry is unstable to
a sliding instability in the presence of vacancies.
This results in the phase near full packing having columnar order, where there is translational order
only in one of the three directions. This phase also has roughly equal number of both types of particles. 
In the presence of attractive interactions between the arms of the particles, 
we argue, using a high density expansion at finite temperatures, that this result continues to hold.
Thus, irrespective of whether attractive interactions are present, neither does  the high density phase have sublattice order nor is there
a critical temperature above which there is no phase transition in contradiction to the results reported in Refs.~\cite{2015-rthrg-tsf-impact,C3RA45342A}.
We also demonstrate the presence of an intermediate phase, and that there are two entropy-driven phase transitions with increasing density of
particles: first from a disordered phase to an intermediate density sublattice phase 
where the symmetry between the two kinds of particles are broken and second from the sublattice phase to a high density columnar
phase where the symmetry between the two types of particles is restored. 
In addition, we also study the special case of the model when only one kind of $Y$-shaped particle is present, and show that it undergoes
a single first order transition from a disordered phase to an ordered sublattice phase.

The remainder of the paper is organized as follows. In Sec.~\ref{model}, we define the model precisely and
explain the algorithm that we use to equilibrate the system in grand canonical Monte
Carlo simulations. The different phases of the model and the nature of the phase transitions for
systems with only one type of particle and both types of 
particles are numerically obtained in Sec.~\ref{onep} and Sec.~\ref{twop} respectively. 
Section.~\ref{conclusion} contains a summary and discussion of the
results.

\section{Model and algorithm \label{model}}
\begin{figure}
\includegraphics[width=\columnwidth]{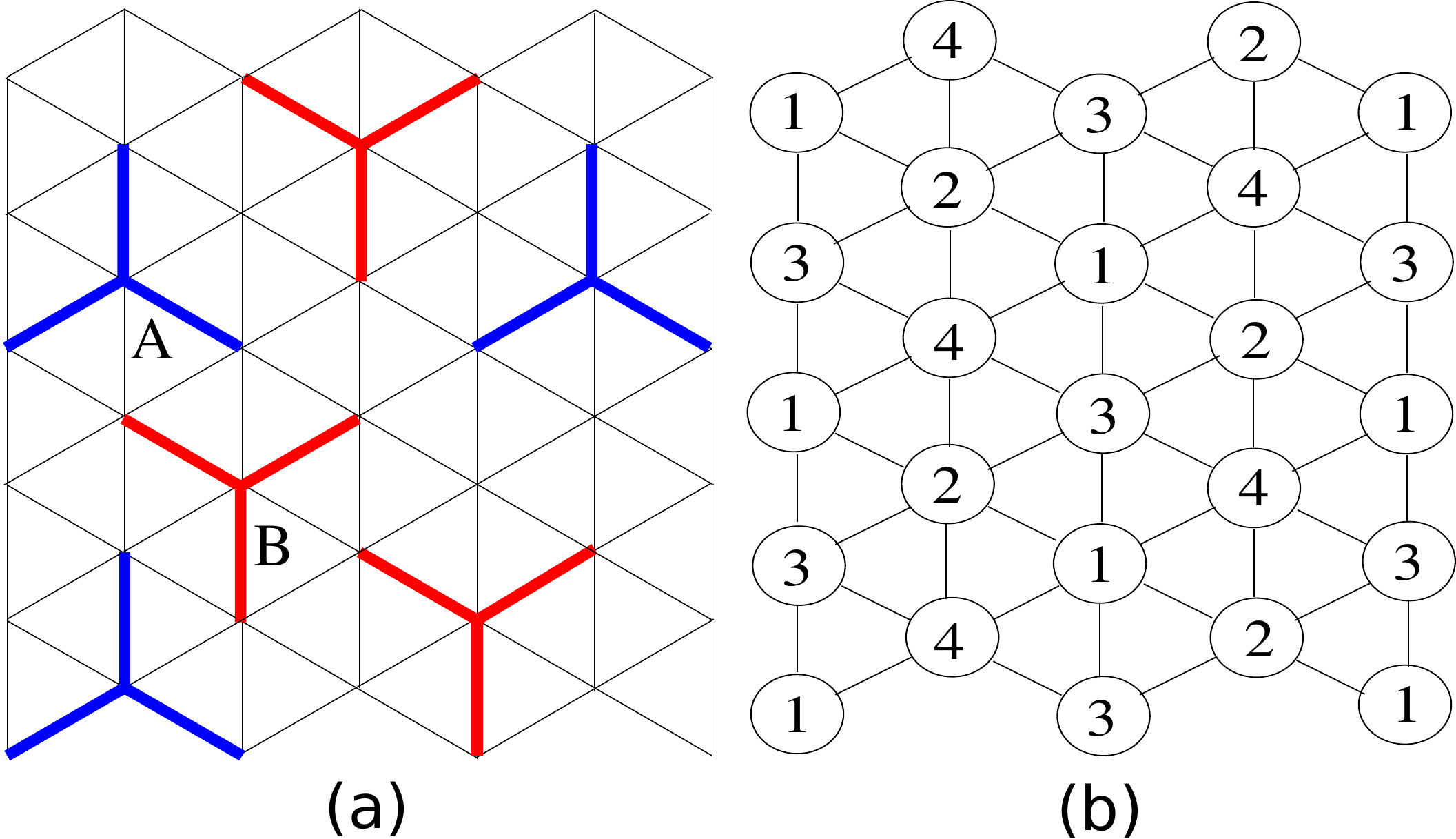}
\caption{(a) Schematic diagram of a triangular lattice and the  two types of $Y$-shaped particles. $A$- and $B$-type 
particles are represented by blue and red colors respectively. (b) The lattice sites are 
labeled as $1, 2,3,4$ depending on the sublattice they belong to.}
\label{particle}
\end{figure}

Consider a two dimensional triangular lattice  of linear dimension $L$ with periodic boundary, as shown in Fig.~\ref{particle}(a). 
A lattice site may be empty or occupied  by one of  two types of particles. Particles are $Y$-shaped and 
occupy four lattice sites, consisting of a central site and three of its six neighbors chosen alternately.   The three neighbors can 
be chosen in two different ways, and hence there are two types of particles, examples of which are
shown in Fig.~\ref{particle}(a). We will refer to the two types as   $A$- and $B$-type particles.
The particles interact through excluded  volume interaction, i.e., a site may be occupied by utmost one particle.
Activities   $z_A=\exp(\mu_A)$ and $z_B=\exp(\mu_B)$ are associated with each $A$- and $B$-type 
particle respectively, where $\mu_A$ and $\mu_B$ are the reduced chemical potentials. 
We will refer to the central site
of a particle as its head.

We study the system using grand canonical Monte Carlo simulations. Conventional algorithms involving local
evaporation and deposition of a single particle  are inefficient in equilibrating the system at densities
close to full packing. We implement an improved version of a recently introduced  algorithm with cluster moves that is able to 
efficiently equilibrate systems of particles with large excluded volume interactions at densities close 
to full packing~\cite{2012-krds-aipcp-monte,2013-krds-pre-nematic}, as well as at the fully packed density~\cite{2015-rdd-prl-columnar}.

We briefly describe the algorithm. First, a row is chosen at random (the row can be in any of the three directions of
the triangular lattice). Then all the $A$-type (or equivalently $B$-type)
particles with heads on this row are evaporated. The row now consists of empty intervals separated from each other by $B$-type particles
with heads on the same row as well as $A$- and $B$-type particles with heads on neighbouring rows. These empty intervals are now re-occupied with
$A$-type particles with the correct equilibrium probabilities. The calculation of these probabilities reduces to determining the 
partition function of a one-dimensional system of dimers. Details may be found in Refs.~\cite{2012-krds-aipcp-monte,2013-krds-pre-nematic,2014-nr-pre-multiple,2015-rdd-prl-columnar}.
For each row, we choose at random whether $A$- or $B$-type particles are to be evaporated. 
A Monte Carlo move is completed when $3 L$ rows are updated.

\begin{figure}
\includegraphics[width=\columnwidth]{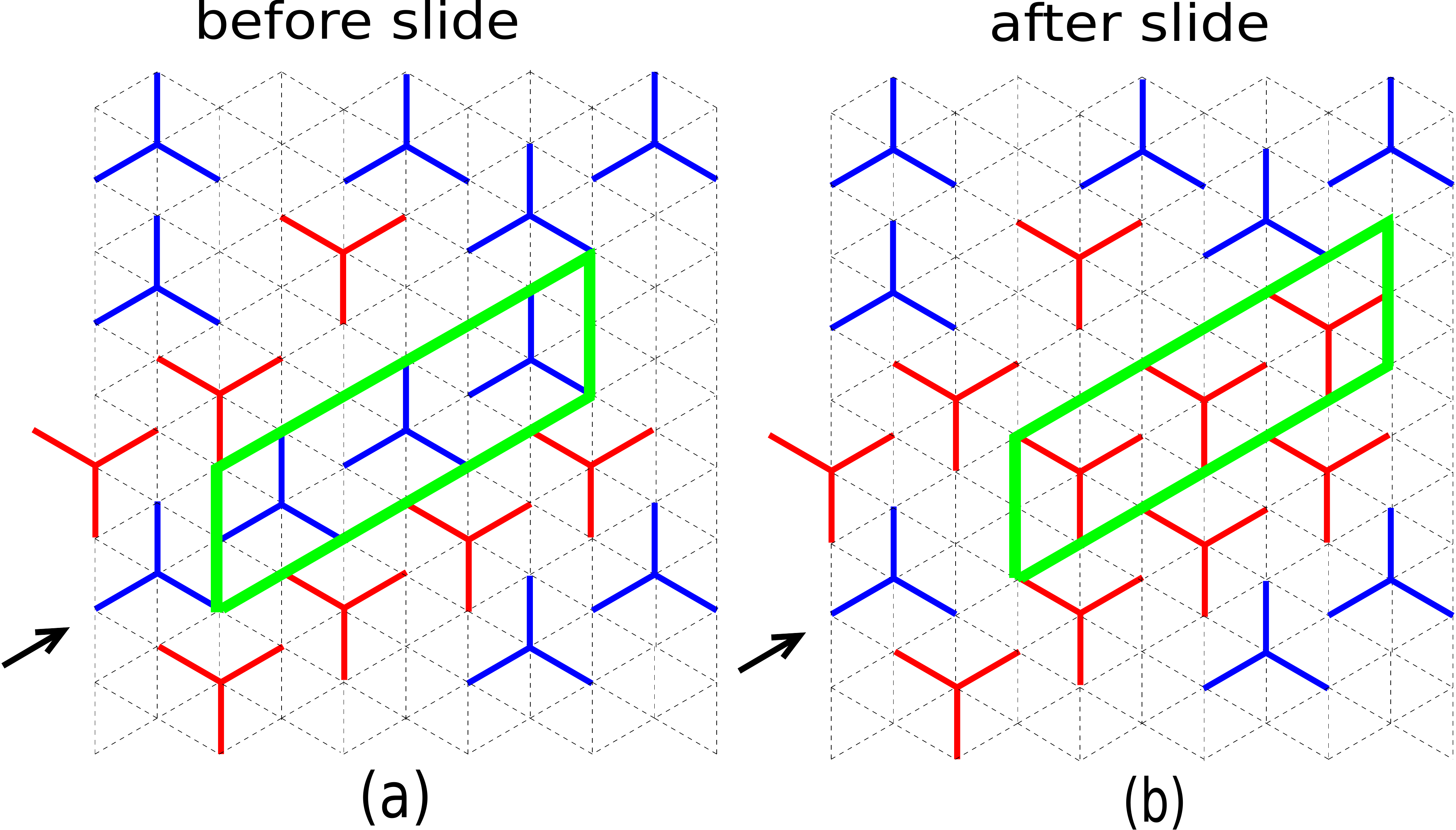}
\caption{Schematic diagram to illustrate the sliding move. A cluster is identified [highlighted box in (a)] by choosing randomly  a site  
and one of the six directions (shown by arrow). The cluster is slid by one lattice site in the chosen direction and the particle type is switched from
$A \leftrightarrow B$ to obtain a new configuration as shown in (b).}
\label{lattice}
\end{figure}

Though the above algorithm is able to equilibrate the system at densities close to full packing, we find that the equilibration times 
as well as the autocorrelation times are large. In order to improve the efficiency of the algorithm, we introduce a sliding  move in addition 
to the evaporation-deposition move. The first step in the sliding move is to select a site at random. 
If the site is not occupied by the head of a particle, then another site is chosen. If the site is occupied by the head of 
a particle, then one direction out of six possible directions is chosen, and
we identify a cluster of same type of particles, defined as a set of consecutive  particles separated by two sites, starting from the randomly 
chosen site along the chosen direction. An example  of such a cluster is shown by  the highlighted box in Fig.~\ref{lattice}(a). The cluster of particles
is slid by one lattice site in the chosen direction and the particle type is changed from $A \leftrightarrow B$ [see Fig.~\ref{lattice}(b)]. 
The new configuration is accepted if it does not violate the hard core constraint. It is straightforward to confirm that the sliding move obeys detailed 
balance as the reverse move occurs with exactly the same probability. 
A Monte Carlo move is completed when $3 L$ rows are updated through the evaporation-deposition
and $L^2/10$ sliding moves are attempted. We have chosen a ratio of sliding to evaporation/deposition moves that is
efficient but have not optimized the ratio.

\begin{figure}
\includegraphics[width=\columnwidth]{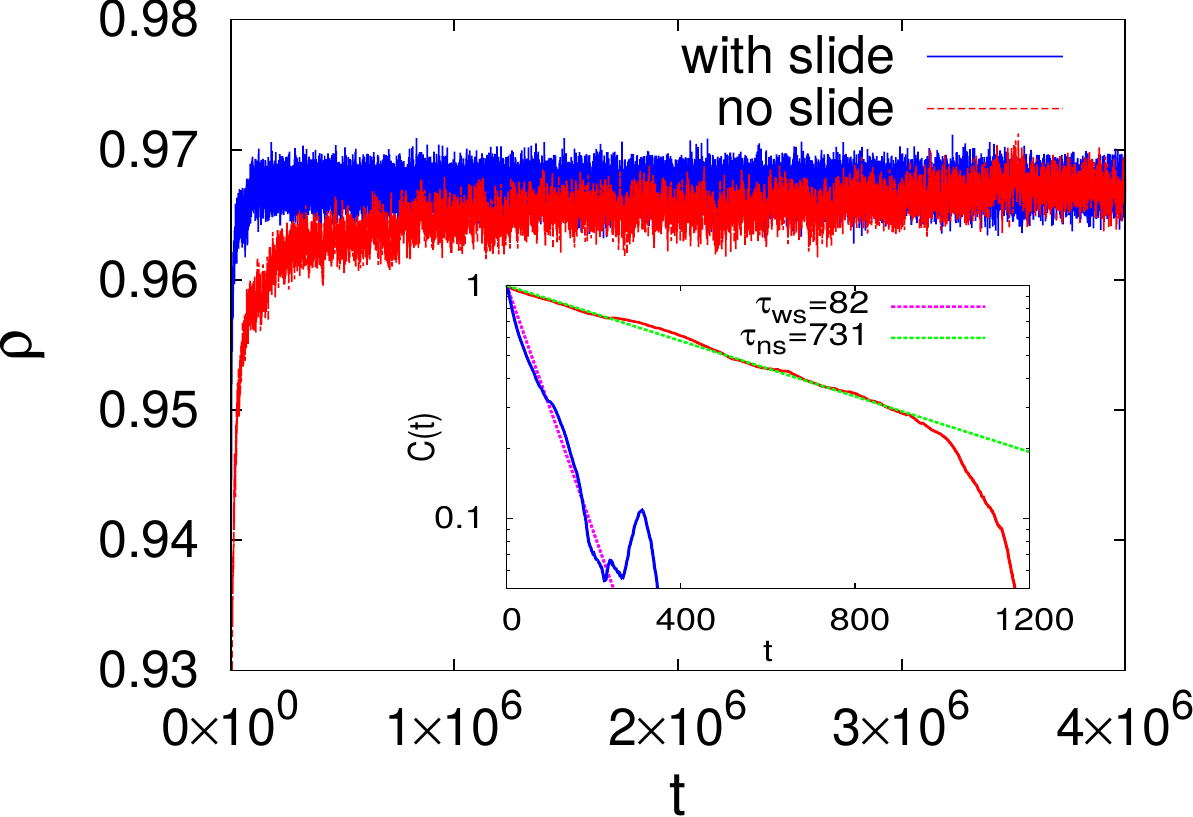}
\caption{The increase in density $\rho$ to its equilibrium value for a system of size $L=300$ and $\mu=\mu_A=\mu_B=6.0$ for the algorithms with (blue) and without (red)
the sliding move. The initial condition is disordered and the equilibrium configuration has $\rho\approx 0.967$, and is ordered.
Inset:  Equilibrium density-density autocorrelation function $C(t)$ as a function of time $t$. When fitted to an exponential as in Eq.~(\ref{eq:11}), we
obtain $\tau_{ws}\approx 82$ when the sliding move is present and $\tau_{ns}\approx 731$ when the sliding
move is absent. }
\label{flip}
\end{figure}

We compare the efficiency
of the algorithm with and without the sliding move in Fig.~\ref{flip}. Starting from
a disordered phase, the system is evolved in time at a value of chemical potential $\mu=\mu_A=\mu_B$
for which the equilibrium density is high ($\approx 0.967$),
and the system is ordered. From Fig.~\ref{flip}, we see that the density reaches
the equilibrium value in $10^5$ steps when the sliding move is present compared
to $4\times 10^6$ steps when the sliding move is absent.
Second, we calculate the
density-density autocorrelation function, defined as 
\be
C(t)=\frac{\langle \rho(t+t_0) \rho(t_0)\rangle-\langle \rho \rangle^2}{\langle \rho^2\rangle-\langle \rho \rangle^2},
\ee
where $\rho(t)$ is the density at time $t$, and the average is over $t_0$. We determine the autocorrelation time $\tau$ by fitting the correlation
function to an exponential
\be
C(t)\approx e^{-t/\tau}.
\label{eq:11}
\ee
From  the inset of Fig.~\ref{flip}, we find the autocorrelation time, $\tau_{ws}$, for the algorithm with sliding move is $\tau_{ws}\approx 82$,
while the  autocorrelation time, $\tau_{ns}$, for the algorithm with no sliding move is $\tau_{ns}\approx 731$.  
Thus, the inclusion of the sliding move results in considerable shorter equilibration times as well as autocorrelation times.

The evaporation and deposition of particles along a row depends only on the configuration of the four neighbouring rows.
Thus, rows that are separated by three can be updated simultaneously, and the implementation of the algorithm
is easily parallelizable. All the results presented in this paper
are obtained using the parallelized algorithm. Equilibration is checked by starting the simulations with different initial conditions,
corresponding to different phases, and confirming that the equilibrated phase  is independent of the initial
condition.

\section{One type of particle ($\bf{z_A=0}$)\label{onep}}

\begin{figure}
{\includegraphics[width=\columnwidth]{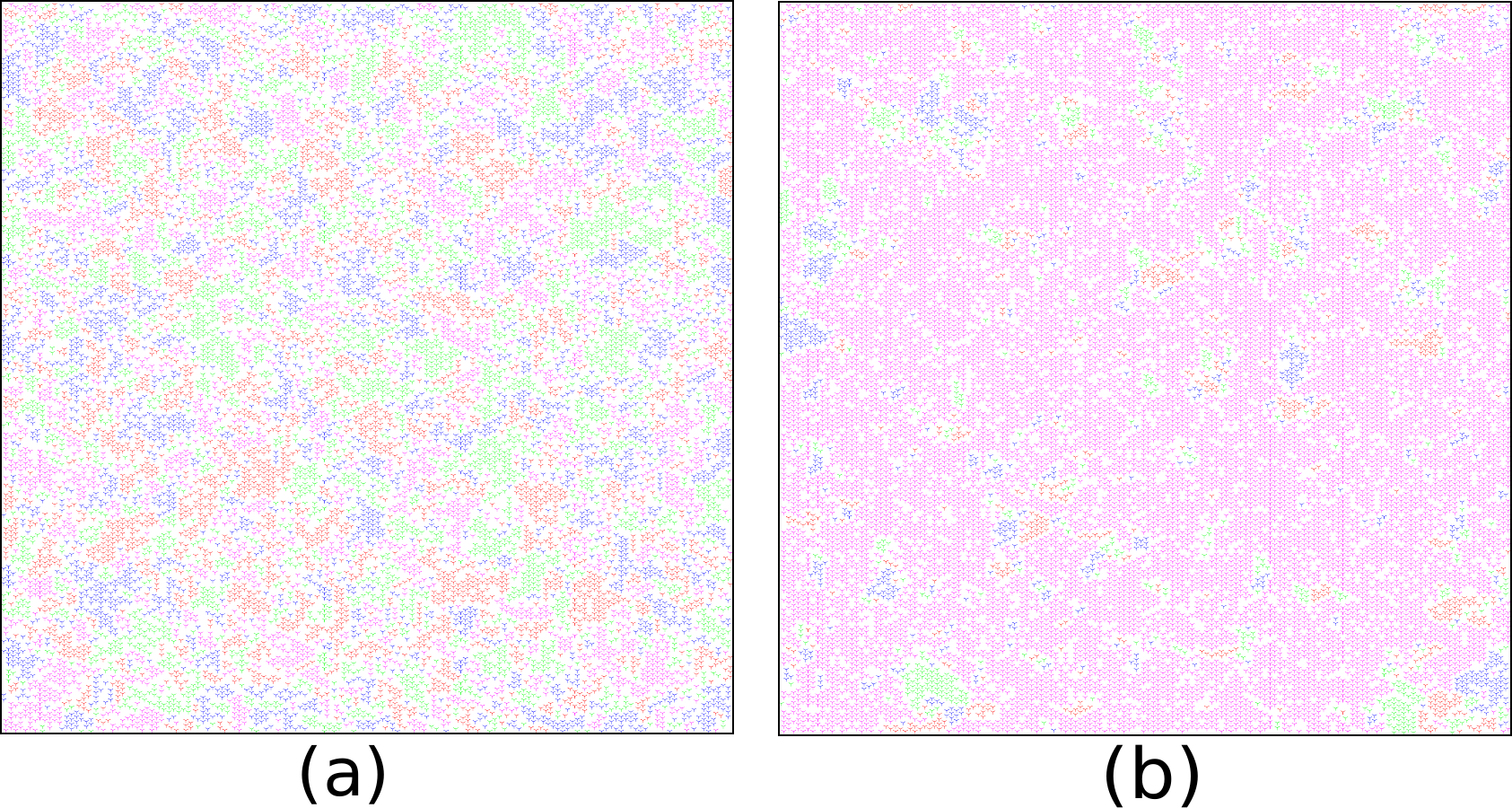}}
\caption{Snapshot of typical equilibrated configurations of the system obtained from {\it{grand canonical}} Monte Carlo simulations 
with only one type of particle ($z_A=0$) for two different values of chemical potential: (a) disordered phase at $\mu_B=1.420$ $(\rho^B \approx 0.710)$ 
and (b) sublattice phase at $\mu_B=1.765$ $(\rho^B \approx 0.775)$. 
The particles on the four sublattices 1, 2, 3 and 4  are represented by green, red, blue and magenta respectively. The data are for a system of size $L=300$.}
\label{snap_onep}
\end{figure}

We first obtain the phase diagram for  the case when only $B$-type particles are present, corresponding  to
$z_A=e^{\mu_A}=0$ and $z_B=e^{\mu_B} > 0$.
To demonstrate the different types of phases present in the system, we divide the  lattice into four sublattices
as shown in Fig.~\ref{particle}(b). A particle occupies four sites that belong to four different sublattices.
We color the four sites occupied by a particle by one of four colors depending
on the sublattice that the head of the particle belongs to. 
Snapshots of typical equilibrated configurations are shown in
Fig.~\ref{snap_onep} for both small densities [Fig.~\ref{snap_onep}(a)] and high densities [Fig.~\ref{snap_onep}(b)].
From the snapshots, it is clear that at small densities, all four colors are roughly equally present. We will refer
to this phase as the disordered phase, in which
\be
\rho_1^B \approx \rho_2^B \approx \rho_3^B \approx\rho_4^B,~~\textrm{disordered phase},
\label{eq:density}
\ee
where $\rho_i^B$ is the fraction of sites in sublattice $i$ that are occupied by $B$-type particles. 

The snapshot of the system at higher densities, as shown in Fig.~\ref{snap_onep}(b) is predominantly of one color, implying that
the heads of the  particles preferably occupy one of the four sublattices.
We will refer to this solid-like  phase as the sublattice phase. The sublattice phase has  translational
order.

To quantify the phase transition from the disordered phase to sublattice phase, we define the vector
\begin{equation}
\label{QB}
{\bf{Q}}^B=|{\bf{Q}}^B| e^{i\theta_B}=\sum_{n=1}^4 \rho_n^B e^{i (n-1)\pi/2},
\end{equation}
where the sublattice densities $\rho_i^B$ are as defined in Eq.~(\ref{eq:density}).
We define the sublattice order parameter $Q^B$ to be
\be
Q^B=\langle |{\bf{Q}}^B|\rangle,
\ee 
where the average $\langle \ldots \rangle$ is over equilibrium configurations. Clearly, $Q^B$ is zero in the
disordered phase and non-zero in the sublattice phase.

\begin{figure}
{\includegraphics[width=\columnwidth]{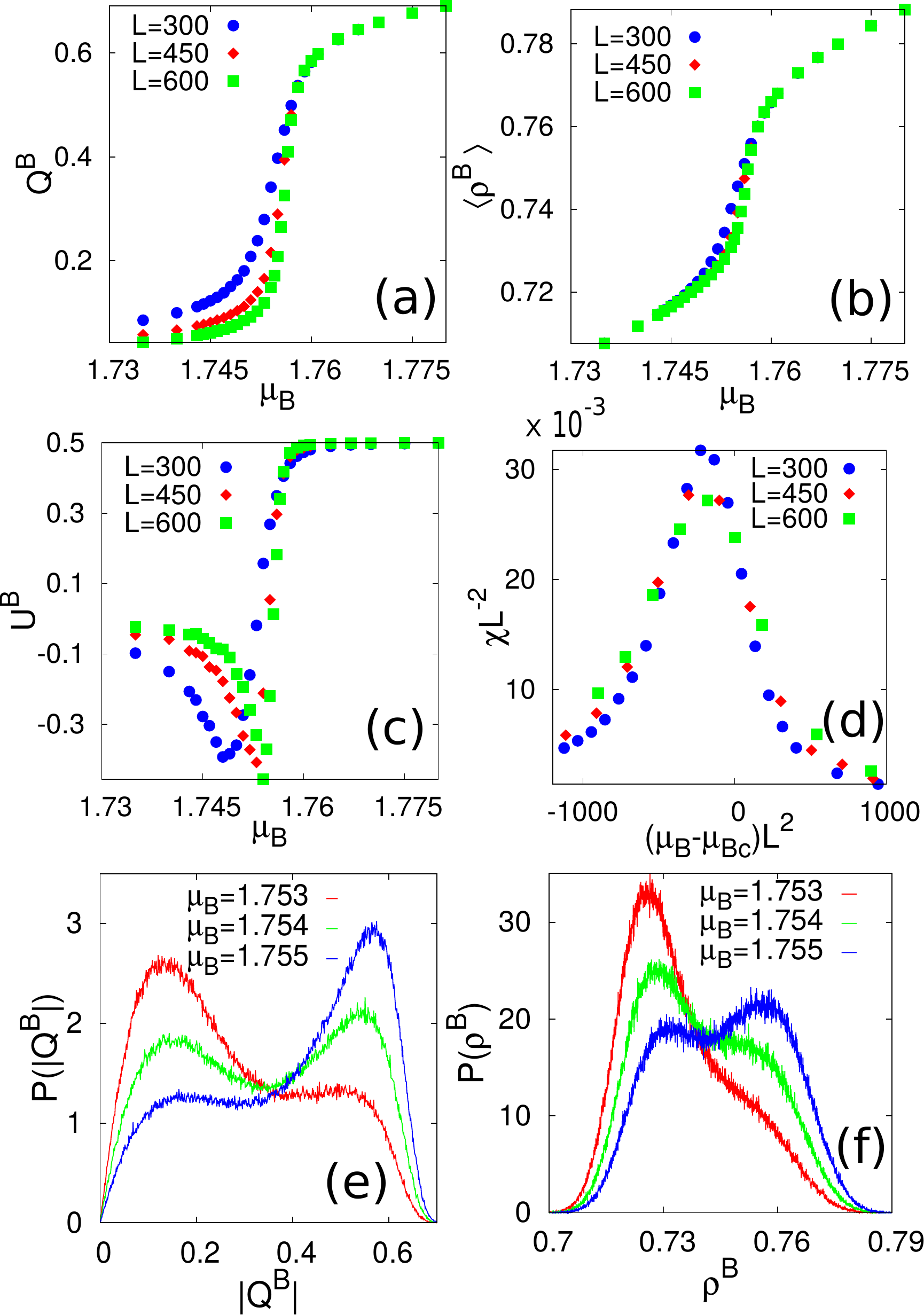}}
\caption{Plot of (a) order parameter $Q^B$, (b) total density $\rho^B$, (c) Binder cumulant $U^B$ and (d) $\chi L^{-2}$, as a function of
chemical potential $\mu_B$ for three different system sizes $L=300$ (blue), $450$ (red), $600$ (green).
Plot of probability density function (e) $P(|\bf{Q^B}|)$, (f) $P(\rho^B)$ for $\mu_B=1.753, 1.754, 1.755$
with system size $L=300$.}
\label{one_particle}
\end{figure}

The variation of $Q^B$ with chemical potential $\mu_B$ is shown in Fig.~\ref{one_particle}(a) for different system sizes.
It increases sharply from zero to a non-zero value as $\mu_B$ crosses a critical value $\mu_{Bc} \approx 1.75$ and critical density $\rho^B_c\approx 0.750$. The transition becomes
sharper with increasing system size. The total density of the system $\rho^B$ has a system size dependence for intermediate
densities [see Fig.~\ref{one_particle}(b)]. We also study the Binder cumulant $U^B$ defined as
\be
U^B=1-\frac{\langle {|{\bf{Q}}^B|}^4 \rangle}{2\langle {|{\bf{Q}}^B|}^2 \rangle^2}.
\ee
The variation of $U^B$ with $\mu_B$ is shown in Fig.~\ref{one_particle}(c)  for three different system sizes. 
For small $\mu_B$, it is zero  for the disordered phase and close to $0.5$ for the ordered phase as expected.
Near the transition point, $U^B$ becomes negative and the minimum value decreases with increasing
system size. This is a clear signature of a first order transition, as for a continuous transition $U^B$ is
positive and the data for different system sizes intersect at the critical point. We conclude that the
transition is first order.
Now, consider the susceptibility $\chi$ defined as
\be
\label{chib}
\chi=L^2(\langle {{|{\bf{Q}}^B|}}^2 \rangle-{Q^B}^2).
\ee
For a first order transition, the singular behaviour of $\chi$ near the transition obeys the
finite size scaling $\chi \sim L^2 f[(\mu_B-\mu_{Bc})L^2]$, where $f$ is a scaling function.
The data for $\chi$ for different system sizes collapse onto one curve when scaled as above with
$\mu_{Bc}\simeq1.756$ as shown in Fig.~\ref{one_particle}(d). 

\begin{figure}
{\includegraphics[width=\columnwidth]{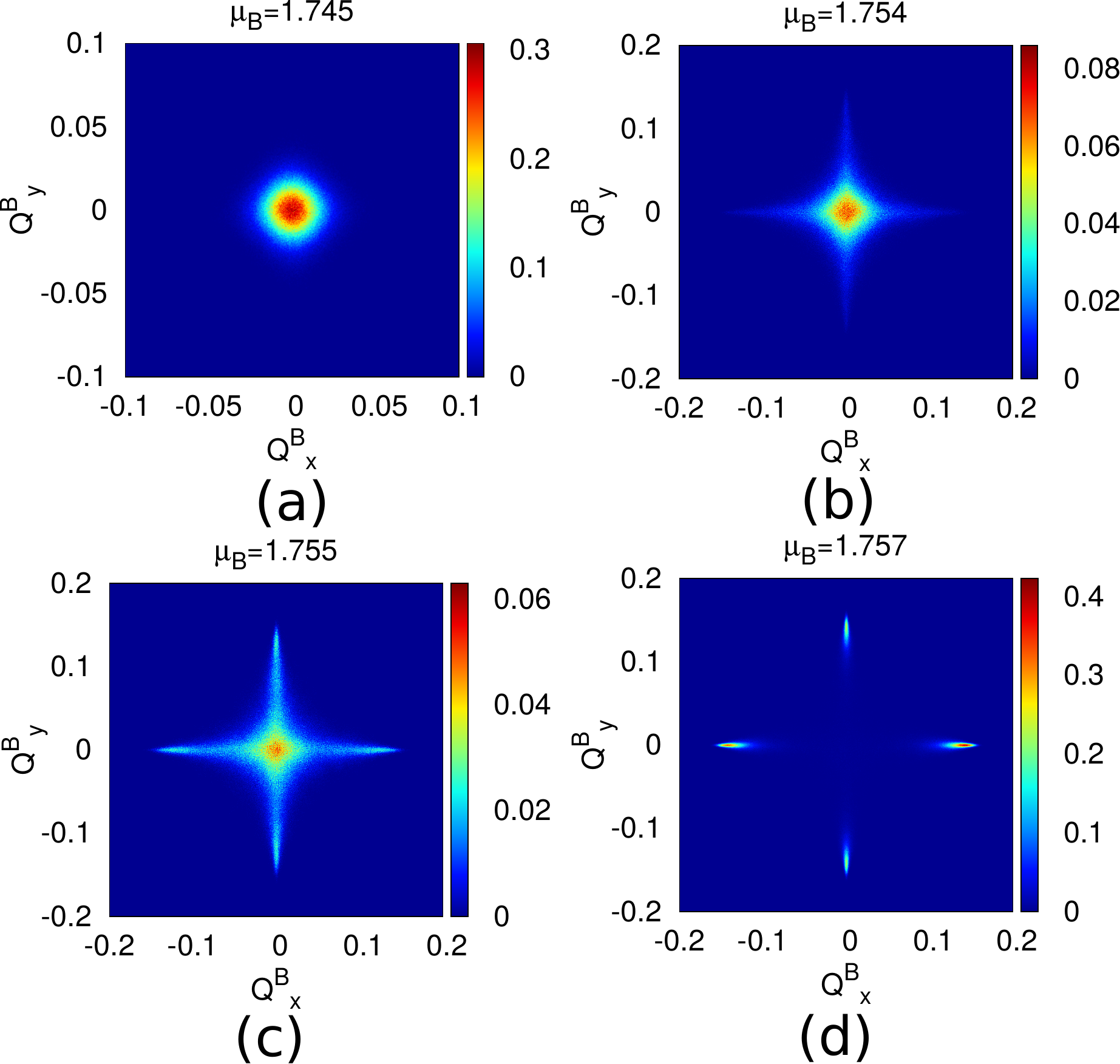}}
\caption{Two dimensional density plots of $P(\bf{Q^B})$ for different values of $\mu_B$ near the transition: (a) $\mu_B=1.745$, (b) $\mu_B=1.754$, (c) $\mu_B=1.755$,
and (d) $\mu_B=1.757$. The data are for a system of size $L=600$.}
\label{colourplot}
\end{figure}

We now give further evidence of the transition being first order.
At a first order phase transition, the system keeps transiting from the disordered phase to the sublattice phase. This 
results in the probability distributions for the order parameter and density having multiple peaks. The probability distribution
for $|\bf{Q^B}|$ and the density $\rho^B$ are shown in Figs.~\ref{one_particle}(e) and (f) respectively for values of $\mu_B$ near the
transition point. The plots shows two 
clear peaks for $\mu_B \approx 1.75$, one corresponding to the disordered phase and the other to the
sublattice phase,  consistent with a first order transition. The two dimensional color plot of the
probability distribution of the complex order parameter
$\bf{Q^B}$ near the critical point is shown in Fig.~\ref{colourplot}, and is consistent with the above observation.

\begin{figure}
{\includegraphics[width=0.8\columnwidth]{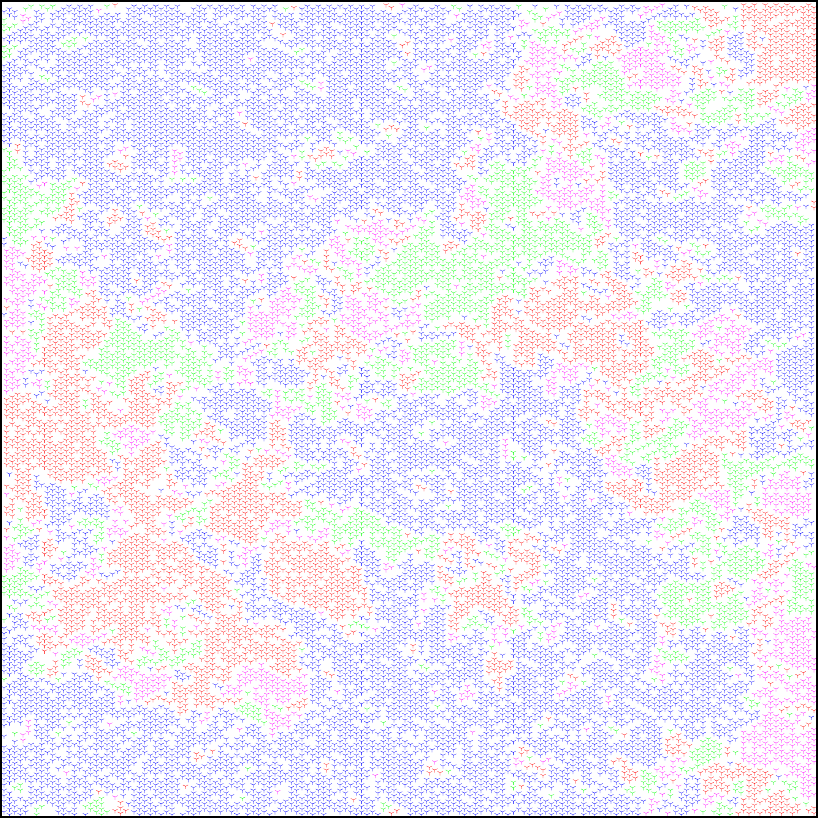}}
\caption{Snapshot of a typical equilibrated configuration of the system obtained from {\it{canonical}} Monte Carlo simulations with one type of particle
having fixed density $\rho^B=0.740$. The particles on the  four sublattices 1, 2, 3 and 4  are represented by green, red, blue and 
magenta respectively. The snapshot shows the co-existence of the sublattice and disordered phases. The data are for a system of size $L=300$.}
\label{snap_mxwl}
\end{figure}

To further establish the first order nature of the transition, we show coexistence of the disordered phase and sublattice phase at the transition
point. To do so, we do simulations in the canonical ensemble, conserving density, of a system having density that lies between
the density of disordered system just below the transition and the density of the sublattice phase just above the transition.
We choose  $\rho^B=0.74$, which lies between the two maxima of the probability distribution for density as shown in  Fig.~\ref{one_particle}(f).
The system is evolved in time through an algorithm that  conserves the density of the system. A lattice site is chosen at random.
If it is occupied by the head of a particle, the particle is removed and deposited at another randomly chosen lattice site. If the deposition
does not violate the hard core constraint, the move is accepted, else the particle is placed at its original position.
The algorithm obeys detailed balance as each move is reversible and occurs at the same rate. 
The snapshot of a typical equilibrated configuration of the system is shown in Fig.~\ref{snap_mxwl}.
There are regions where the colour is uniform (blue), showing a sublattice phase, while there are other regions where all four colours appears, corresponding
to a disordered phase.  We conclude that there is phase segregation and coexistence, both signatures of a first order transition.

\section{Two types of Particles ($\bf{z_A=z_B}$)\label{twop}}

Now consider the case where both types of particles are present with equal activity $z_A=z_B=z$. 
It is natural to expect that the fully packed phase has a sublattice order where the heads of particles
occupy only one sublattice. We first argue that 
at densities close to full packing, sublattice order is not stable due to the presence of vacancies, and the system prefers a columnar
order with densities of both types of particles being roughly equal. We illustrate this instability through an example. 

\begin{figure}
\includegraphics[width=\columnwidth]{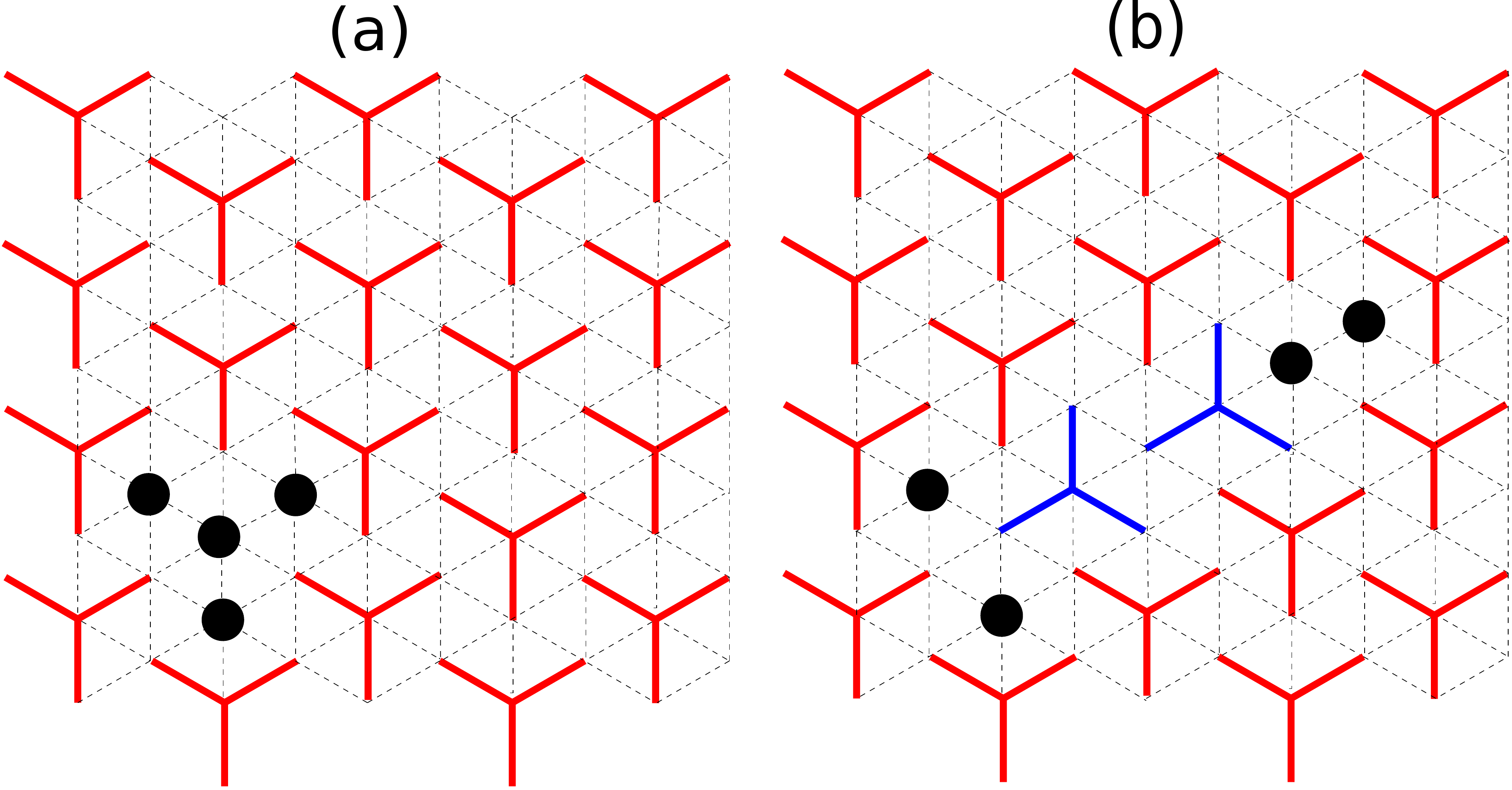}
\caption{(a) Schematic diagram showing the creation of a vacancy consisting of four empty sites (black solid circles), when a particle is removed from
the fully packed sublattice phase. (b) The vacancy may be split into two half-vacancies, and separated along a row 
by sliding particles along the row and changing the type.}
\label{s_to_c}
\end{figure}

Consider a fully packed configuration with sublattice order. Such a configuration can have only one type of particle (say B-type).
Removal of a single particle creates single vacancy made of four empty sites as shown by the filled circles in Fig.~\ref{s_to_c}(a). 
These empty sites may be split into two unbound pairs of half-vacancies by sliding a number of consecutive particles adjacent to the empty sites
and flipping their type to A, each of these configurations having the same weight.  
An example of two particles being slid is shown in Fig.~\ref{s_to_c}(b). Introducing more vacancies results
in destabilizing the sublattice phase. Sliding results in restoring translational invariance along two of the three directions. However,
translational order is still present in the third direction. We will refer to this 
phase as the columnar phase. We note that in this phase, 
two sublattices are preferentially occupied, one with $A$-type particles and the other with $B$-type particles. 
The stabilization of the columnar phase
by creating vacancies is an example of order by disorder, prototypical example being the hard square 
gas~\cite{1967-bn-jcp-phase,1966-bn-prl-phase,1966-rc-jcp-phase,2012-rd-pre-high,2016-ndr-epl-stability,2017-mnr-jsm-estimating}.

If additional attractive interactions are present between neighbouring arms, then the above argument may also be extended to account for 
the energy cost of creating vacancies. It may then be shown that even for this case the columnar phase is preferred over the 
sublattice phase. To preserve continuity of presentation,
we postpone describing the generalized argument to Sec.~\ref{conclusion}.

\begin{figure}
{\includegraphics[width=\columnwidth]{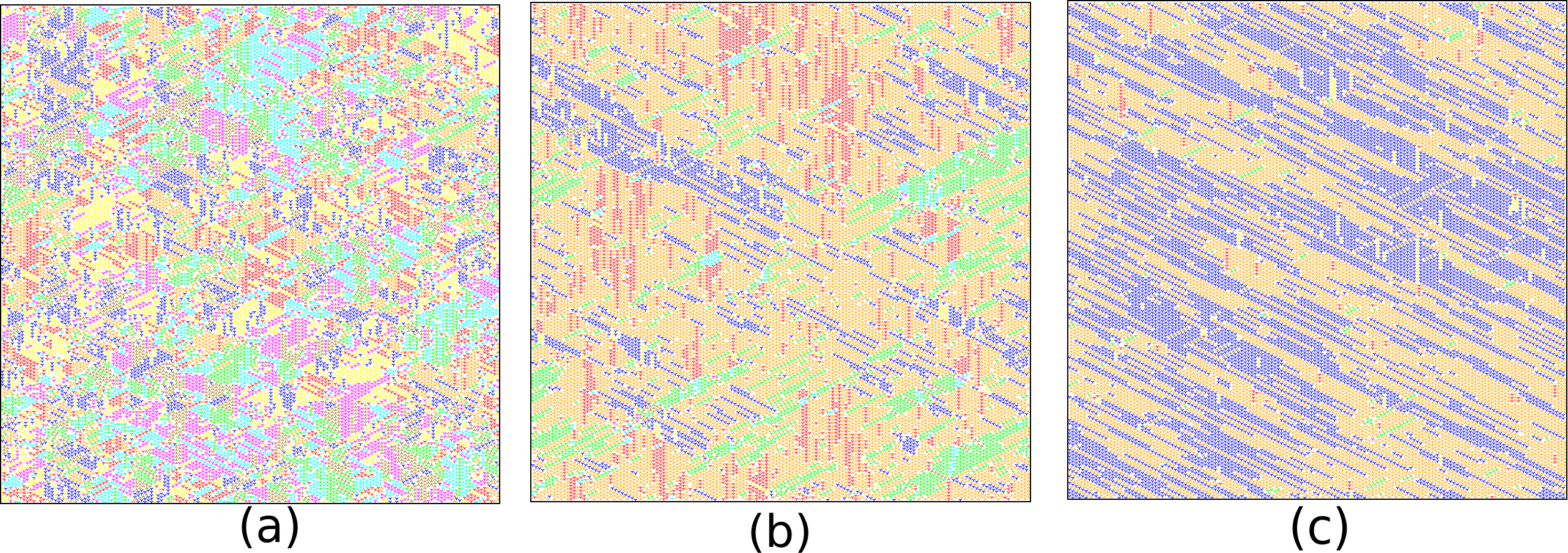}}
\caption{Snapshots of equilibrated configurations of the system with two types of particles obtained from {\it grand canonical} Monte Carlo simulations 
for different values of $\mu$: (a) disordered phase with $\mu=4.5$ ($\rho \approx 0.88 $), (b) sublattice phase with $\mu=5.4$ ($\rho \approx 0.947$),
and (c) columnar phase with $\mu=6.0$ ($\rho \approx 0.967$). 
The particles on the four sublattices 1, 2, 3 and 4  are represented by  yellow, olive, cyan and orange for type A and
by  green, red, blue and magenta for type B. The data are for a system of size $L=300$.}
\label{phases_twop}
\end{figure}

We now give numerical evidence for the high density phase being columnar and also determine numerically the different phases of the system 
at densities away from full packing.
Snapshots of equilibrated configurations of the system for different values of $\mu$ are shown in Fig.~\ref{phases_twop}. Here, the lattice
sites are colored using eight colors depending on the type of the particle (2 types) and the sublattice (4 sublattices)
that the head belongs to. For small values
of $\mu$, the snapshot contains all eight colours distributed uniformly [see Fig.~\ref{phases_twop}(a)], corresponding to the disordered
phase. For intermediate values of $\mu$, the snapshot shown in Fig.~\ref{phases_twop}(b) is predominantly of one color. This
phase corresponds to a sublattice phase. The sublattice phase breaks the A-B symmetry and one type of particle is preferred over the
other. Finally, for larger values of $\mu$, the snapshot shown in Fig.~\ref{phases_twop}(c) has mostly two colors that appear
in strips. This phase corresponds to the columnar phase.
This is in agreement with our argument presented above that the phase close to full packing
is columnar due to the sublattice phase being unstable due to a sliding instability. We thus identify two phase transitions, the critical values of $\mu$
being denoted as $\mu^{DS}$ and $\mu^{SC}$.

The sublattice phase has an 8 fold degeneracy. To quantify it, consider the vector ${\bf{Q}}_s$:
\be
{\bf{Q}}_s=|{\bf{Q}}^A|-|{\bf{Q}}^B|,
\ee
where ${\bf{Q}}^B$ is given in Eq.(\ref{QB}) and ${\bf{Q}}^A$ has a similar definition with
$\rho_n^B$ replaced by $\rho_n^A$.
We define the sublattice order parameter $Q_s$ to be
\be
\label{eq:q_s}
Q_s=\langle |{\bf{Q}}_s|\rangle.
\ee 
$Q_s$ is zero in the disordered phase and non-zero in the sublattice phase. It is also straightforward to check that $Q_s\approx 0$ in the
columnar phase. 
We characterize the fluctuations of $Q_s$ through the susceptibility $\chi_s$ defined as
\be 
\chi_s=L^2(\langle |{\bf{Q}_s}|^2 \rangle-Q_s^2).
\ee
We also define the Binder cumulant associated with $Q_s$ as $U_s$:
\be
\label{eq:us}
U_s=1-\frac{\langle |{\bf{Q}}_s|^4 \rangle}{2\langle |{\bf{Q}}_s|^2 \rangle^2}.
\ee

To characterize the symmetry breaking between the two types of the particles in the disordered phase, we
introduce an order parameter $\rho_d$ defined as 
\be
\label{eq:rho_d}
\rho_d=\langle |\rho^A - \rho^B| \rangle,
\ee
where $\rho^A$ and $\rho^B$ are the fraction of sites occupied by $A$ and $B$-type particles respectively.
We denote the associated susceptibility as $\chi_d$ and Binder cumulant as $U_d$:
\bea 
\chi_d&=&L^2\left[\langle (\rho^A - \rho^B)^2 \rangle-\rho_d^2\right],\\
U_d&=&1-\frac{\langle |\rho^A - \rho^B|^4 \rangle}{2\langle |\rho^A - \rho^B|^2 \rangle^2}.
\label{eq:ud}
\eea
\begin{figure}
{\includegraphics[width=\columnwidth]{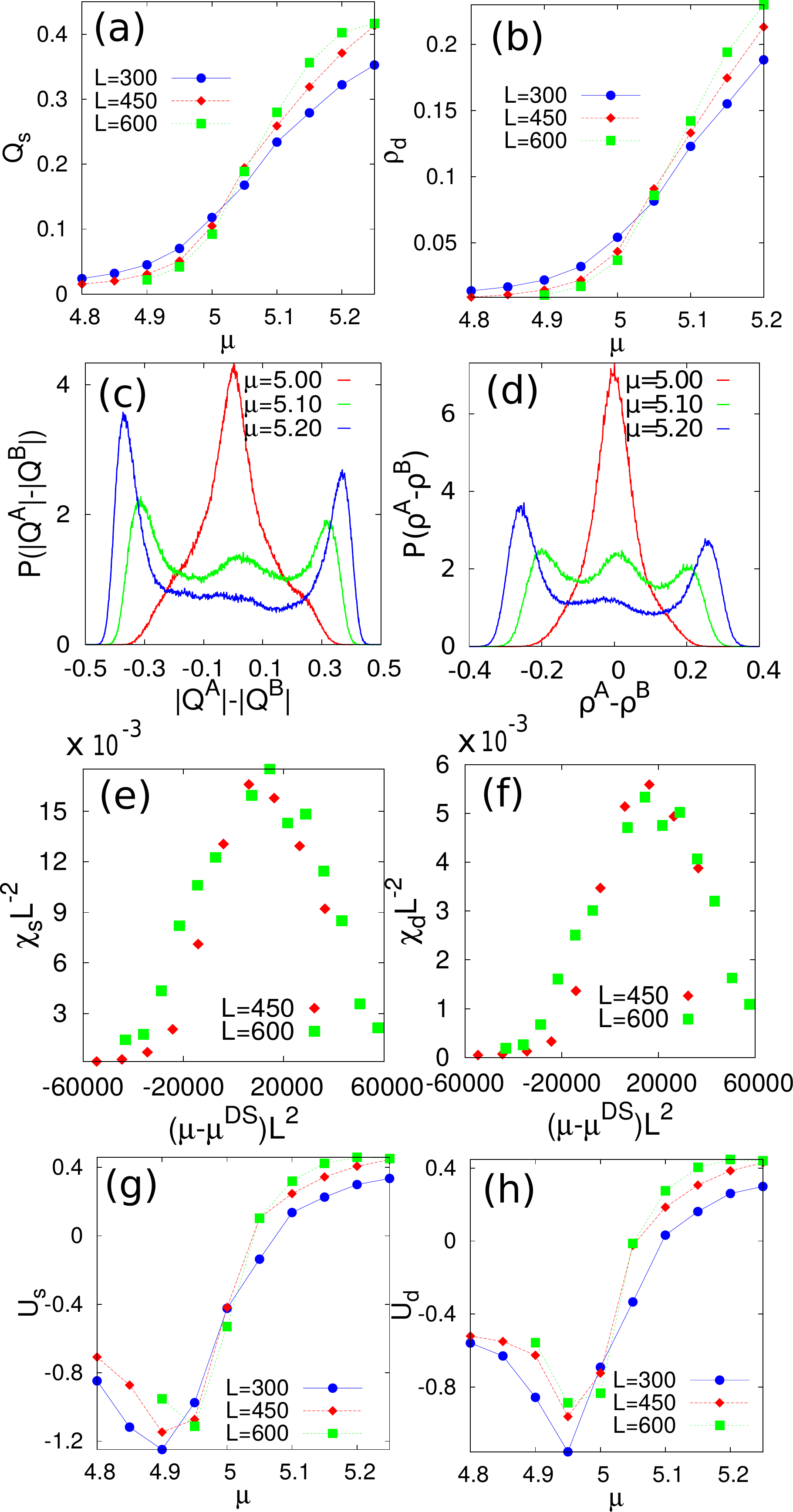}}
\caption{Plot of (a) sublattice order parameter $Q_s$ and (b) density difference $\rho_d$ as a function of $\mu$.
Plot of probability distribution: (c) $P(|Q^A|-|Q^B|)$ and (d) $P(\rho^A-\rho^B)$ near disorder 
to sublattice transition for the system of size $L=300$.
Plot of rescaled susceptibilities: (e) $\chi_sL^{-2}$ and (f) $\chi_dL^{-2}$ associated with $Q_s$ and $\rho_d$ respectively about the critical
point $\mu^{DS}$. Plots are for the systems of size $L=300, 450$ and $600$. 
}
\label{s_trans}
\end{figure}

The variation of the order parameters $Q_s$ and $\rho_d$ with $\mu$ is shown in Fig.~\ref{s_trans}(a) and (b) respectively.  They increase from
close to zero to a nonzero value, showing the presence of the sublattice phase. The curves for different system sizes cross close
to $\mu \approx 5.07$, and density $\rho \approx$ 0.930. While a clear discontinuity in the order parameters is not 
discernable from Fig.~\ref{s_trans}(a) and (b), we now present evidence for the transition being first order in nature. The probability distributions
for ${\bf{Q}}_s$ and $|\rho^A - \rho^B|$ near the transition point are shown in Fig.~\ref{s_trans}(c) and (d) respectively. As $\mu$ is increased,
the probability distributions change from being single peaked, corresponding to the disordered phase, to a three-peaked distribution,
corresponding to coexistence of the sublattice and disordered phase, to a symmetric double-peaked distribution, corresponding to
the sublattice phase. Coexistence close to the transition is a clear signature of the first order nature of the transition. We note that the
distributions sharpen with increasing system size. 
The variation of Binder cumulant $U_s$ and $U_d$ with $\mu$ is shown in Fig.~\ref{s_trans}(g) and (h) respectively. It becomes negative
for certain values of $\mu$, which is a clear signature of first order transition.
In a first order transition, the susceptibilities scale as
\be
\label{eq:scaling}
\chi \sim L^2f[(\mu-\mu_c)L^2].
\ee
When scaled as described with $\mu^{DS}\approx5.07$, $\rho^{DS}\approx0.930$ , the data for 
different system sizes collapse onto a single curve, as shown in Fig.~\ref{s_trans}(e) and (f). We conclude that the disordered to sublattice
transition is first order in nature.

In the columnar phase, two sublattices are preferentially occupied by the particles. This selection can be done in six different ways and each way
has two possibilities of filling, as $A$-type and $B$-type particles can choose either one of the selected two sublattices.
Thus, the columnar phase has a $12$ fold degeneracy.
To quantify this phase illustrated in Fig.~\ref{phases_twop}(c), we define a columnar order parameter
$Q_c$ as follows. In the columnar phase, the particles occupy alternate rows along one of the three orientations,
and occupy all rows in the other two orientations. The breaking of the translational invariance in a direction
is reflected in the difference in density of heads between even and odd rows and is captured by 
\bea
Q_1 &=& |\rho_1+\rho_2-\rho_3 -\rho_4|, \nonumber\\
Q_2 &=& |\rho_1 + \rho_3-\rho_2 -\rho_4|, \label{eq:qi}\\
Q_3 &=& |\rho_1 + \rho_4-\rho_3 -\rho_2|, \nonumber
\eea
where $\rho_i$ is the fraction of sites belonging to sublattice $i$ that is occupied by a particle, irrespective of the type.
In $Q_1$, $(\rho_1 + \rho_2)$ measures the density of occupied sites in odd horizontal rows [see Fig.~\ref{particle}(b)] and $(\rho_3 + \rho_4)$
the density of occupied sites in even horizontal rows. Thus, $Q_1$ is non-zero only when there is translational order along the
horizontal rows, and similar interpretations hold for $Q_2$ and $Q_3$. Now, consider the vector
\be
{\bf Q}_c=|{\bf Q}_c |e^{i\theta_c}=Q_1+Q_2e^{2\pi i/3}+Q_3e^{4\pi i/3}.
\ee
We define the columnar order parameter to be
\be
Q_c=\langle |{\bf Q}_c|\rangle.
\ee
In the columnar phase, $Q_c$ is non-zero. In the disordered phase $Q_c \approx 0$, as each of the $Q_i$ in Eq.~(\ref{eq:qi}) is approximately zero.
In the sublattice phase, one sublattice is preferentially occupied and each of the $Q_i$ in Eq.~(\ref{eq:qi}) becomes non-zero but approximately
equal in magnitude, and hence again $Q_c \approx 0$. Thus, a non-zero $Q_c$ is a signature for the columnar phase. We define
the corresponding susceptibility as 
\be
\chi_c=L^2(\langle |{\bf{Q}}_c|^2 \rangle-Q_c^2).
\ee
\begin{figure}
{\includegraphics[width=\columnwidth]{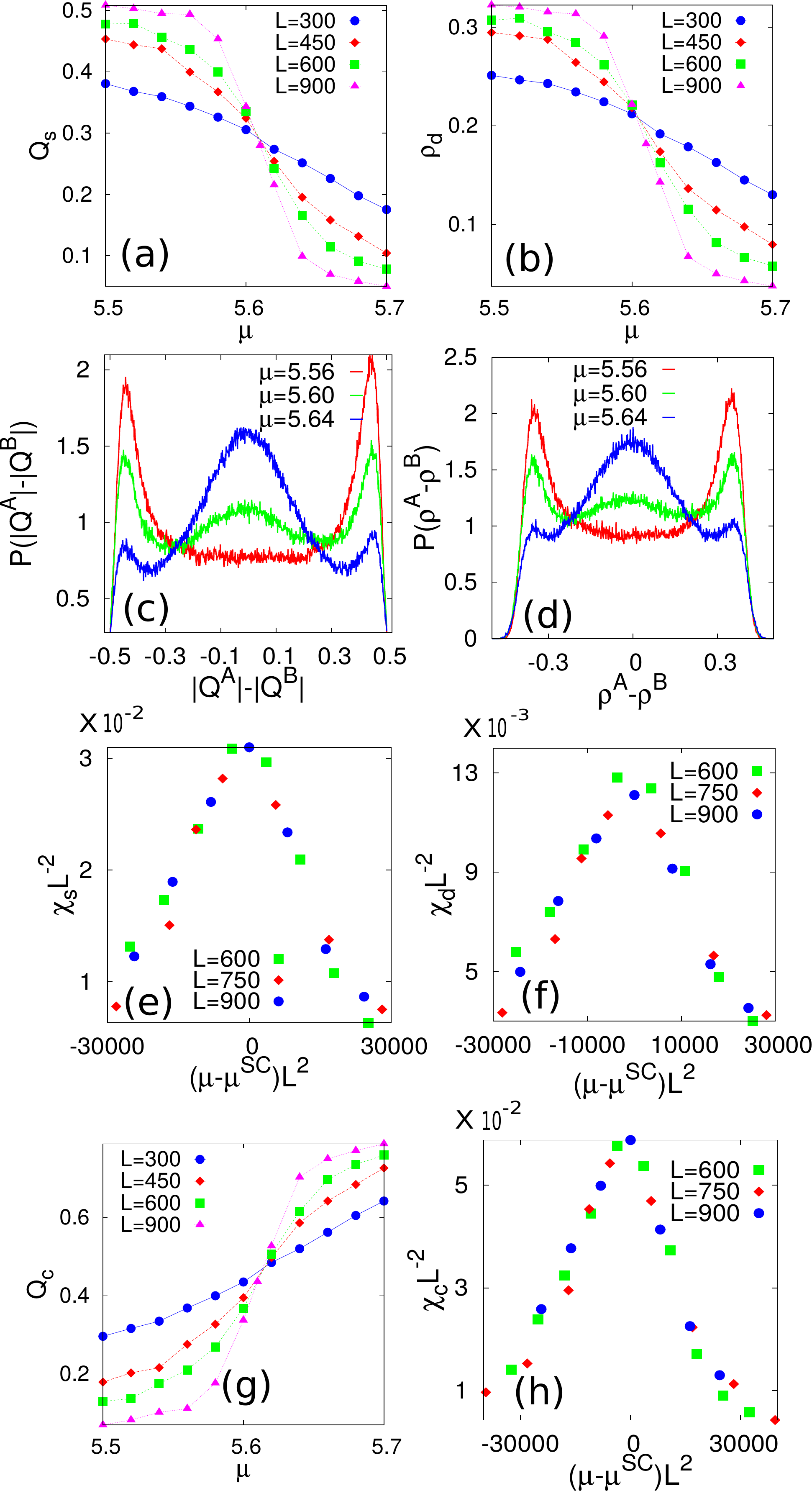}}\label{col_trans}
\caption{Plot of (a) sublattice order parameter $Q_s$ and (b) density difference $\rho_d$ as a function of $\mu$.
Plot of probability distribution: (c) $P(|Q^A|-|Q^B|)$  and (d) $P(\rho^A-\rho^B)$  near
sublattice to columnar transition for the system size $L=300$.
Plot of rescaled susceptibilities: (e) $\chi_sL^{-2}$ and (f) $\chi_dL^{-2}$ associated with $Q_s$ and $\rho_d$ respectively about the critical
point $\mu^{SC}$.
Plot of (g) columnar order parameter $Q_c$ and (h) associated rescaled susceptibility $\chi_cL^{-2}$ as a function of $\mu$ and $(\mu-\mu^{SC})$ respectively.
Plots are for the systems of size varying from $L=300$ to $900$.
}
\label{c_trans}
\end{figure}

In the columnar phase the sublattice order parameter $Q_s$ [see Eq.~(\ref{eq:q_s})] and the density difference $\rho_d$ [see Eq.~(\ref{eq:rho_d})] both becomes zero.
The variation of $Q_s$ and $\rho_d$ with $\mu$ is shown in Fig.~\ref{c_trans}(a) and (b) respectively.
The probability distributions for $Q_s$ and $(\rho^A - \rho^B)$ near the
transition point are shown in Fig.~\ref{c_trans}(c) and (d) respectively.
As $\mu$ is increased, the probability distributions
change from symmetric double-peaked, corresponding to the sublattice
phase, to a three-peaked distribution, corresponding to coexistence of the sublattice and columnar
phase, to a dominant single-peaked  distribution, corresponding to the columnar phase. The coexistence
of both columnar phase and sublattice phase is a signature of first order transition. 
For the first order transition susceptibilities follow the scaling law as described in Eq.~(\ref{eq:scaling}).
With this scaling we get the collapse of susceptibilities $\chi_s$ and $\chi_d$ onto single curve as shown 
in Fig.~\ref{c_trans}(e) and (f), for critical value of chemical potential $\mu^{SC}\approx5.61$
with density $\rho^{SC}\approx0.956$.

The variation of the order parameter $Q_c$ with $\mu$ for different system size is shown in Fig.~\ref{c_trans}(g). It 
acquires nonzero value in the columnar phase. The susceptibility $\chi_c$ also obeys the scaling law as described in
Eq.~(\ref{eq:scaling}). This is confirmed from the Fig.~\ref{c_trans}(h) in which collapse of curves for different
system sizes with described scaling is shown.

\section{Conclusion\label{conclusion}}

In this paper we studied the different phases and  phase transitions of hard $Y$-shaped particles on a two dimensional triangular lattice. 
There are two types of $Y$-shaped particles depending on their orientation on the lattice, which are mirror images of
each other. By incorporating cluster moves, we were able to equilibrate the system at densities close to full packing, 
allowing us to unambiguously determine  the phases at all densities. In addition to the low-density disordered phase, we find
two other phases. At intermediate phases, the phase has a solid-like sublattice order. In this phase, the symmetry between 
the two types of particles is broken resulting in a majority of one type of particle. In addition, these particles preferentially occupy
one of the four sublattices of the lattice. At high densities, the phase has columnar order. In this phase, the symmetry between the
two types of particles is restored. However, there is translational order in one of the three directions, wherein particles preferentially
occupy either even or odd rows. 
The first transition from disordered to sublattice phase occurs
at $\mu^{DS}\approx5.07$  and the second transition from sublattice to columnar phase occurs at $\mu^{SC}\approx5.61$. 
Both the transitions are first order in nature. $Y$-shaped particles give a simple example of a system where small number of vacancies
destabilize the sublattice phase into a columnar phase while a larger number of vacancies again stabilizes the sublattice phase.
When only one type of particle is present, the model undergoes a single first order phase transition from a
low density disordered phase to high density sublattice phase, and occurs at $\mu_{Bc}\approx1.756$.

\begin{figure}
\includegraphics[width=\columnwidth]{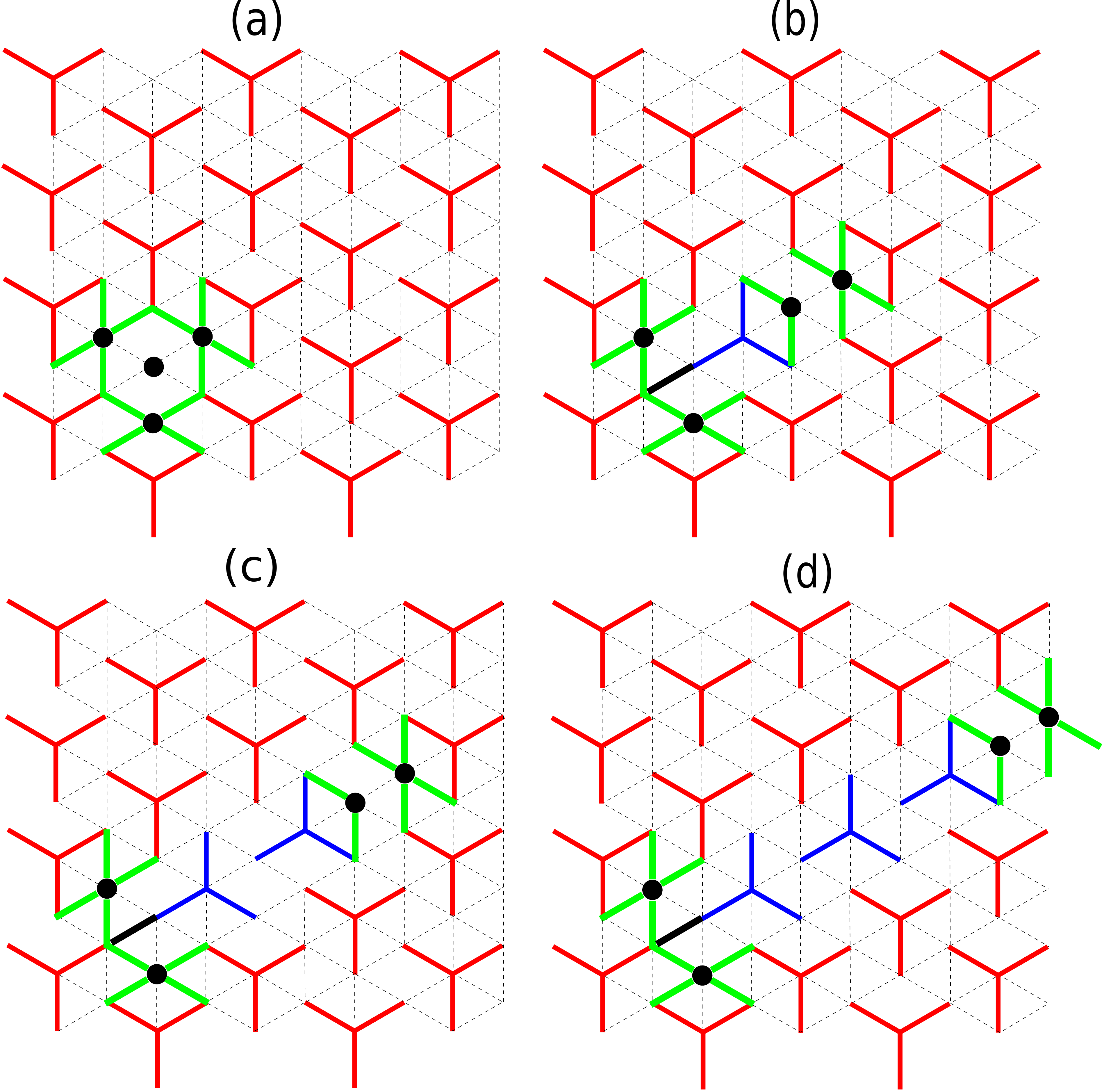}
\caption{Schematic diagrams to calculate, in a fully packed sublattice phase of $B$-type particles, the energy cost to create: 
(a) a vacancy consisting of four empty sites (black solid circle) vacancy), 
(b) two half-vacancies separated  by one $A$-type particle,
(c) two half-vacancies separated by two $A$-type particles, and
(d) two half-vacancies separating three $A$-type particles.
Compared to the background sublattice phase, green bonds increase the energy by $\epsilon$ while black bonds decrease the energy by $\epsilon$. 
The energy cost is $12 \epsilon$ for (a), and $13 \epsilon$ for (b)-(d).
}
\label{armlength1}
\end{figure}

The high density phase that we observe in this paper has columnar order with both types of particles equally present,  which is in contradiction to the  
results obtained from Monte Carlo simulations of $Y$-shaped particles with attractive interactions in Refs.~\cite{2015-rthrg-tsf-impact,C3RA45342A}, wherein 
it was shown that the high density phase 
has sublattice order in which only one kind of particle is present. For only excluded volume interactions, we argued in Sec.~\ref{twop}
that the introduction of vacancies 
results in the destabilization of the sublattice phase, 
because the vacancies split into two unbound half-vacancies that can be separated without any cost in entropy. We now argue that this instability is present
even in the presence of
attractive interaction between the nearest-neighbour arms of different particles. Consider the case when vacancy is created by removing a single
particle from a sublattice phase at full packing, as shown in Fig.~\ref{armlength1}(a). If $-\epsilon$ is the energy of each nearest neighbour pair of arms,
then this vacancy costs an energy $12 \epsilon$. On splitting the vacancy into two half-vacancies and sliding them away from each other by one, two,
or three particles [see Figs.~\ref{armlength1}(b)--(d)], the energy cost increases to $13 \epsilon$, but {\it does not} increase with separation between
the half-vacancies. Thus, the partition function $Z$ of the system may be written as
\bea
Z&=&4 z^{N/4} e^{3 N\beta \epsilon/2} \bigg[ 1+ \frac{N e^{-12\beta \epsilon}}{4 z} \nonumber\\
&&+ \frac{3 N (\frac{L}{2}-1)e^{-13\beta \epsilon}}{8 z} +\mathcal{O}(z^{-2})\bigg],
\eea
where $\beta=(k T)^{-1}$ is the inverse temperature.
The free energy $\beta f =-\ln Z$ is then
\be
\beta f= \frac{-\ln z}{4} - \frac{3 \beta \epsilon}{2}-\frac{e^{-12 \beta \epsilon}}{4 z}- \frac{3 L e^{-13 \beta \epsilon}}{16 z}+\mathcal{O}(z^{-2}).
\label{eq:arm1}
\ee
Clearly, the  term proportional to $z^{-1}$ diverges with system size, as is indicative of systems with columnar order. If the divergent terms are
resummed correctly, taking into account the columnar nature of the phase, then the first correction term becomes $\mathcal{O}(z^{-1/2})$~\cite{2012-rd-pre-high}. The divergent term shows that the expansion about the sublattice phase is not convergent and, thus, we conclude that the high density phase is columnar even when interactions are present. We note that in Ref.~\cite{C3RA45342A}, attractive
interactions were included for neighbouring central sites too. However, it may be easily checked that the above expansion is true for this case also,
albeit with a energy cost of $2 \epsilon$ for  half-vacancy when compared to bound vacancy. From Eq.~(\ref{eq:arm1}), it may also be seen that for temperatures less than or order of $\epsilon/\ln L$, it would be possible to see a sublattice phase, but this is purely a finite-size effect.

In addition, it was argued in Ref.~\cite{2015-rthrg-tsf-impact,C3RA45342A} that there is no phase transition above a critical temperature. However, the results in this paper correspond to the limit of infinite temperature, wherein we established the presence of two transition. Re-analysing the model with interactions to make the results consistent
with those in this paper is a promising area for future study. Another area for future study is 
the system of $Y$-shaped particles with larger arm lengths which could be symmetric~\cite{C3RA45342A}. 
For these systems with only excluded volume interaction, we expect the high density phase to 
be columnar~\cite{C3RA45342A}.

It is tempting to analyse the high density columnar phase using high density expansions as developed for squares and 
rectangles~\cite{1967-bn-jcp-phase,1966-bn-prl-phase,1966-rc-jcp-phase,2012-rd-pre-high,2016-ndr-epl-stability,2017-mnr-jsm-estimating,2015-rdd-prl-columnar,2017-mr-pre-columnar}. These expansions are in terms of number  defects (which could be extended).
However the columnar phase of $Y$-shaped particles is different from that of these simpler models, in which the type of particles occupy preferred sublattices
in the columnar phase. This makes it difficult to even write the zeroth order term for the partition function corresponding to no defects.

HCLGs sometimes show multiple phase transitions with increasing density, but only when the excluded volume per particle is large. For instance, for 
multiple phase transitions to be present, the minimum range of interaction is seventh nearest neighbour for 
rods~\cite{2013-krds-pre-nematic,2017-vdr-arxiv-different}, fifth nearest neighbour for rectangles~\cite{2014-kr-pre-phase,2015-kr-pre-asymptotic},
fourth nearest neighbour for HCLG models for discs~\cite{2014-nr-pre-multiple,2016-nr-jsm-high} while nearest neighbour exclusion models like
the $1$-NN model  on the square 
lattice~\cite{1967-bn-jcp-phase,1988-ps-jsp-classical,1980-bl-prb-phase,2007-fal-jcp-monte,1965-gf-jcp-hard,1980-bet-jsp-hard,1994-bf-jcp-hard,1965-r-prl-hard,1966-rc-jcp-exact,1966-rc-jcp-phase,1974-nf-physica-hard,2002-gb-pre-finite,2012-c-jpa-series,2012-j-jpa-comment,1983-m-jsp-monte,1989-hm-prb-percolation,2000-le-prb-ordering,2003-lc-pre-density,1980-r-prb-phase,1991-hc-prb-percolation}
or the hard hexagon model on the triangular lattice~\cite{1980-b-jpa-exact}
show only one transition from a disordered phase to a sublattice phase. The excluded volume of $Y$-shaped particles consists of nearest neighbour 
sites, as
in the hard hexagon model
and half of the next-nearest neighbour sites depending on the pair of particles being considered. It is quite surprising that despite the
short ranged nature of the interaction, the system undergoes two density-driven phase transitions. 
It is possible that this feature may also be extended to mixtures on a square lattice. From the insights gained from the current paper, 
we expect that 
if there are two kinds of particles $A$ and $B$ on a square lattice, where the $A$-$A$ and $B$-$B$ excluded volume interactions are upto second nearest neighbour, but the
$A$-$B$ excluded volume interaction is upto the third nearest neighbour, then the high density phase will be columnar and there will be multiple transitions.
Confirming this conjecture in simulations would be interesting.

\section*{Acknowledgements}
The simulations are carried out in the supercomputing machines Annapurna and Nandadevi at The Institute of Mathematical Sciences. We thank 
Deepak Dhar, J\"urgen Stilck, and Kedar Damle for helpful discussions.

%

\end{document}